\documentclass[prd,twocolumn,nofootinbib]{revtex4}
\usepackage{graphicx, epsfig}
\usepackage{color}
\usepackage{mathrsfs}
\usepackage{bm}
\usepackage{amsmath,amssymb}
\usepackage[caption=false]{subfig}
\usepackage{hyperref}
\usepackage[normalem]{ulem}
\usepackage{mathtools}
\usepackage{subfig}

\newcommand{\be}{\begin{equation}}
\newcommand{\ee}{\end{equation}}
\newcommand{\ba}{\begin{eqnarray}}
\newcommand{\ea}{\end{eqnarray}}
\newcommand{\nn}{\nonumber}

\begin{document}
\title{Quantum Evaporation of Classical Breathers}
\author{Jan Oll\'e$^*$, Oriol Pujol\`as$^*$, Tanmay Vachaspati$^\dag$, 
George Zahariade$^{\dag\sharp}$}
\affiliation{
$^*$Institut de F\'isica d'Altes Energies, Universitat Aut\`onoma de Barcelona, 
E-08193 Bellaterra, Spain \\
$^\dag$Physics Department, Arizona State University, Tempe, AZ 85287, USA. \\
$^\sharp$Beyond Center for Fundamental Concepts in Science, Arizona State University, 
Tempe, AZ 85287, USA. \\
}

\begin{abstract}
\noindent
We apply the recently discovered classical-quantum correspondence (CQC) to study 
the quantum evaporation of breathers in an extended sine-Gordon model.
We present numerical results for the decay rate of the breather as a function of the coupling
strength in the model.
This is a complete treatment of the backreaction 
of quantum radiation on the classical dynamics of oscillons.
\end{abstract}

\maketitle

\section{Introduction}

Oscillons are long-lived localized oscillating solutions of non-linear wave equations. They occur in many different fields of physics, such as condensed matter theory~\cite{cite-key} or cosmology~\cite{Amin:2011hj}, and have been studied extensively. A particularly interesting example of such a solution is the so-called breather solution of the sine-Gordon model~\cite{Ablowitz:1973fn}. It is well-known that this model possesses soliton and antisoliton solutions interpolating between two consecutive vacua, which are non-perturbative, stable and stationary. In this context breather solutions can be interpreted as soliton-antisoliton bound state solutions which are time-periodic and perfectly stable.

It is of course understood that breathers are purely classical solutions but their stability properties can be shown to extend into the quantum realm~\cite{Coleman:1985rnk}. This however ceases to hold when one introduces a coupling to a quantum radiative field. The quantum evaporation of breathers can be studied in the limit where the breather itself is treated classically and provides a time-dependent background causing the radiation field to get excited, thereby losing energy and slowly decaying. The classical breather thus becomes a ``quantum oscillon''. This radiative phenomenon is analogous to particle production during gravitational collapse (Hawking radiation)~\cite{Hawking:1974sw} or pair creation in electric fields (Schwinger pair production)~\cite{Schwinger:1951nm}.

This radiative phenomenon has been studied in detail by Hertzberg in Ref.~\cite{Hertzberg:2010yz} 
but there the backreaction of the quantum radiation on the classical breather background was not fully 
taken into account. In this work we provide a full treatment of the classical breather decay under the 
effect of quantum radiation {\it i.e.} fully incorporating backreaction effects. We do this by using the 
recently developed classical-quantum correspondence (CQC) described at length in Refs.~\cite{Vachaspati:2018llo,Vachaspati:2018hcu,Vachaspati:2018pps}.
 The CQC method is identical to the ``mode function method'' used previously in the literature 
 (see {\it e.g.},~\cite{Aarts:1999zn,Borsanyi:2007wm,Borsanyi:2008eu,Saffin:2014yka}).
 In~\cite{Aarts:1999zn} the technique was used to study fermion production and
 backreaction in classical scalar plus gauge field backgrounds;
 in~\cite{Borsanyi:2007wm,Borsanyi:2008eu,Saffin:2014yka} particle production and backreaction in solitonic backgrounds were treated but also including other assumptions because of particle interactions.

Quantum backreaction on classical backgrounds is generally calculated in the semiclassical
approximation. This is an iterative procedure and usually only the first iteration is carried out
as continuing the iterations is quite laborious. Within the CQC however, the quantum fields
are replaced by corresponding complexified classical variables. The dynamics of these
classical variables together with the classical background defines a new entirely classical
dynamics problem that can be solved by numerical or other means. The resulting solution
for the classical background is precisely the backreacted solution we are seeking.
Application of this technique to a quantum mechanical example, where the full solution can 
also be calculated, shows excellent agreement~\cite{Vachaspati:2018llo}. 

In Sec.~\ref{model} we start by introducing the notations pertaining to the particular 
evaporating breather model we will be studying. The $1+1$ dimensional model will 
involve a classical sine-Gordon field $\phi(t,x)$ and a quantum massless scalar field 
$\psi(t,x)$ coupled together via a bi-quadratic interaction term. In Sec.~\ref{latticeversion} 
we discretize this mixed classical-quantum model on a lattice and map it to a fully classical 
system with well-defined initial conditions that exactly mimics the full backreacted dynamics. 
This mapping is achieved via the CQC which we briefly describe in this particular context. 
In Secs.~\ref{numericalsetup} and \ref{results} we set up the numerical simulation of the dynamics 
and describe its results. We conclude with Sec.~\ref{discussion} where we discuss the validity 
of our analysis and compare it to previous work.

\section{Field theory model}
\label{model}

We consider the sine-Gordon model plus an extra real scalar field with bi-quadratic coupling:
\ba
L= \int dx\Biggr[\frac{1}{2} \dot{\phi}^2&-&\frac{1}{2} \phi'^2 - m_\phi^2 (1-\cos\phi)\nn\\
&+& \frac{1}{2} \dot{\psi}^2- \frac{1}{2} \psi'^2- \frac{\lambda}{2} \phi^2 \psi^2 \Biggr]\,.
\label{fieldaction}
\ea
Here dotted ({\it resp.} primed) quantities denote time ({\it resp.} spatial) derivatives, $m_\phi$
is the mass of the excitations about the vacua, and we use natural units where $\hbar=c=1$.
The classical breather solution,
\be
\phi_b (t,x)= 
4\, \tan^{-1} \left [ \frac{\eta \sin(\omega t)}{\cosh(\eta \omega x)} \right ], \ \ \psi=0
\label{breather}
\ee
with $\eta = \sqrt{m_\phi^2-\omega^2}/\omega$,
is an exact classical solution to the equations of motion that is time-dependent, periodic and 
non-dissipative {\it i.e.} it has an infinite lifetime.
However at the quantum level, the $\psi$ field will get excited and this
quantum radiation will backreact on the breather and cause it to evaporate. 

In general, it is difficult to couple the quantum excitations of the field $\psi$ to the purely classical 
breather solution. Indeed, calculating such backreaction effects requires in principle working in the 
fully quantized theory and since the breather is a non-perturbative solution of the sine-Gordon 
model, results can only be 
obtained via computationally intensive lattice field theory simulations. Another avenue generally
used is the semi-classical approximation whereby the dynamics of the quantum field $\psi$ are 
first determined in the presence of the fixed classical background $\phi$, then the classical equations 
of motion for the field $\phi$ are solved by substituting $\psi^2$ for its vacuum expectation value 
$\langle\psi^2\rangle$, and finally the procedure is reiterated in order to get better and better 
approximations to the backreaction.

In the following, we will choose a somewhat different path and study the evaporation of the breather 
using the classical-quantum correspondence (CQC) developed in 
Refs.~\cite{Vachaspati:2018pps,Vachaspati:2018llo,Vachaspati:2018hcu}. 
Notice that the CQC only applies to the case of free quantum fields. To generalize
the method to scenarios where the field $\psi$ has self-interactions requires approximation
methods~\cite{Borsanyi:2007wm}.

\section{Lattice version}
\label{latticeversion}

We first discretize the theory by putting it on a spatial lattice as in Ref.~\cite{Vachaspati:2018hcu}. More precisely, we introduce an IR regulator $L$ for the spatial domain (physical size of the lattice) and divide the interval $[-L/2,L/2]$ into $N+1$ intervals of size $a=L/(N+1)$. For any integer $i$ running from $0$ to $N+1$ we then define
\ba
\phi(t,ia-L/2)&=&\phi_i (t)\,,\\
\psi(t,ia-L/2)&=&\psi_i(t)
\ea 
and impose Dirichlet boundary conditions $\phi_0=\phi_{N+1}=\psi_0=\psi_{N+1}=0$ at ``spatial infinity." With these definitions \eqref{fieldaction} is well approximated by
\ba
L^{(N)}= \sum_{i=1}^N a\Biggr[\!\!\!\!\!\!\!\!\! &&\frac{1}{2} \dot{\phi}_i^2+\frac{1}{2a^2} \phi_i(\phi_{i-1}-2\phi_i+\phi_{i+1})\nn\\
&+& \frac{1}{2} \dot{\psi}_i^2 + \frac{1}{2a^2}\psi_i(\psi_{i-1}-2\psi_i+\psi_{i+1})\nn\\
&-& m_\phi^2 (1-\cos\phi_i) - \frac{\lambda}{2} \phi_i^2 \psi_i^2\Biggr]\,.
\label{discaction}
\ea
Introducing the matrix $\Omega^2$ defined by
\be 
\Omega^2_{ij}=
\begin{cases}
+2/a^2+\lambda\phi_i^2\,,\ \text{if}\ i=j\\
-1/a^2\,,\ \text{if}\ i=j\pm 1 \,.
\end{cases}
\label{Omega2}
\ee
we can recast the $\psi$ dependent part of \eqref{discaction} in more compact form as
\be
L^{(N)}_\psi = \sum_{i,j=1}^N\left[ \frac{a}{2}\dot{\psi}_i\delta_{ij}\dot{\psi}_j-\frac{a}{2}\psi_i\Omega^2_{ij}\psi_j\right]\,.
\label{psiaction}
\ee
As we have discussed in the previous section, the breather solution \eqref{breather} is classically non-disssipative, 
but not quantumly. In order to study its quantum evaporation, we make use of the techniques developed in Ref.~\cite{Vachaspati:2018hcu}. In the spirit of the CQC, we assume that the $\phi_i$s are classical degrees of freedom while the $\psi_i$s are quantum. At time $t=0$ the initial conditions for $\phi_i$ are such that if $\lambda$ were to be set to zero, one would recover the non-dissipative breather solution. The quantum harmonic oscillators $\psi_i$ on the other hand are taken to be in their ground state initially. 
(Since $\phi(t=0, x)=0$ this corresponds to the quantum field $\psi$ being in its non-interacting ($\lambda=0$) vacuum.) Because of the explicit time-dependence induced by the presence of the $\lambda \phi_i^2$ term in $\Omega^2$, their quantum state subsequently evolves and their average energy increases. Within the CQC, this situation can be accurately and quantitatively described  by promoting the quantum degree of freedom $\psi_i$ to a classical complex $N\times N$ matrix coefficient $Z_{ij}/a$. This results in the following substitution at the level of the discretized Lagrangian:
\be
L^{(N)}_\psi\rightarrow L^{(N)}_Z = \sum_{i,j,k=1}^N \left [ \frac{1}{2a} {\dot Z}^*_{ij}\delta_{ik} {\dot Z}_{kj} - 
\frac{1}{2a} Z^*_{ij} \Omega^2_{ik} Z_{kj} \right ]\,.
\ee
In addition to the above substitution, the CQC paradigm
also requires that we choose very specific initial conditions for the classical variables $Z_{ij}$. These are most easily written in matrix notation:
\be
Z_0=-i\sqrt{\frac{a}{2}}\sqrt{\Omega_0}^{-1}\quad\text{and}\quad\dot{Z}_0=\sqrt{\frac{a}{2}}\sqrt{\Omega_0}\,.
\label{initZ}
\ee 
Here the zero subscript denotes initial values while the square root of a symmetric positive definite matrix $S=O.\text{Diag}(\lambda_1,\lambda_2,\dots,\lambda_N).O^T$ is defined to be $\sqrt{S}=O.\text{Diag}(\sqrt{\lambda_1},\sqrt{\lambda_2},\dots,\sqrt{\lambda_N}).O^T$. 

Finally the CQC Lagrangian reads
\ba
L^{(N)}_{CQC}&=& \sum_{i=1}^N a\Biggr[\frac{1}{2} \dot{\phi}_i^2+\frac{1}{2a^2} \phi_i(\phi_{i-1}-2\phi_i+\phi_{i+1})\nn\\
&&\quad\quad\quad\quad- m_\phi^2 (1-\cos\phi_i) \Biggr]\nn\\
&+&\sum_{i,j,k=1}^N \left [ \frac{1}{2a} {\dot Z}^*_{ij}\delta_{ik} {\dot Z}_{kj} - 
\frac{1}{2a} Z^*_{ij} \Omega^2_{ik} Z_{kj} \right ]\,.
\label{CQCaction}
\ea
Note that the interaction term is present in the very last term because the definition of 
$\Omega^2$ contains $\phi$ as in~\eqref{Omega2}.

In the next section we will set the problem up for numerical simulation.

\section{Numerical setup}
\label{numericalsetup}

To understand how the quantum evaporation of the classical breather proceeds in time we will solve the discretized equations of motion stemming from \eqref{CQCaction}. The evolution of $\phi$ will be described by the ordinary differential equations
\ba
&&\ddot{\phi}_i-\frac{1}{a^2}\left(\phi_{i+1}-2\phi_i+\phi_{i-1}\right)+m_\phi^2\sin(\phi_i)\nonumber\\
&&\hspace{0.5in}+\lambda\left(\frac{1}{a^2}\sum_{j=1}^NZ^*_{ij}Z_{ij}\right)\phi_i=0\label{eqback}\,,
\ea
with initial conditions that would yield the non-dissipative breather solution in the $\lambda=0$ limit {\it i.e.}
\ba
\phi_i(t=0)&=&0\,,
\label{initphiexp}\\
\dot{\phi}_i(t=0)&=& \frac{4\eta\omega}{\cosh\left(\eta\omega a\left(i-\frac{N+1}{2}\right)\right)}\,.
\label{initphidotexp}
\ea
The $Z_{ij}$ will evolve according to
\be
\ddot{Z}_{ij}+\Omega^2_{ik}Z_{kj}=0\,,
\label{eqZ}
\ee
with initial conditions prescribed by the CQC. More precisely, since $\phi_i(t=0)=0$, it is easy to diagonalize 
the tri-diagonal matrix $\Omega_0^2$ and obtain explicit expressions for the initial conditions for $Z_{ij}$. 
Indeed we find that $\Omega_0^2=ODO^{T}$,
where 
\be
D_{ij}=\frac{4}{a^2}\sin^2\left(\frac{\pi i}{2(N+1)}\right)\delta_{ij}\label{approx2}\,,
\ee
and $O$ is an orthogonal matrix with components,
\be
O_{ij}=\sqrt{\frac{2}{N+1}}\sin\left(\frac{\pi ij}{N+1}\right)\label{approx3}\,.
\ee
Therefore using \eqref{initZ} we can explicitly write
\ba
Z_{ij}(t=0)&=&\frac{-ia}{N+1}\sum_{k=1}^N \frac{\sin\left(\frac{\pi i k}{N+1}\right)\sin\left(\frac{\pi k j}{N+1}\right)}{\sqrt{\sin\left(\frac{\pi k}{2(N+1)}\right)}}\,,
\label{initZexp}\\
\dot{Z}_{ij}(t=0)&=&\frac{2}{N+1}\sum_{k=1}^N\sin\left(\frac{\pi i k}{N+1}\right)\sin\left(\frac{\pi k j}{N+1}\right)\nn\\
&&\quad\quad\quad\quad\times\sqrt{\sin\left(\frac{\pi k}{2(N+1)}\right)}\,.
\label{initZdotexp}
\ea
The numerical solution to the coupled system of ordinary differential equations \eqref{eqback} and \eqref{eqZ} with respective initial conditions \eqref{initphiexp},\eqref{initphidotexp}, \eqref{initZexp} and \eqref{initZdotexp} will thus yield the dynamics of the quantum evaporation of the classical breather. In the following we will take $m_\phi=1$, $\omega=0.25$ and $L=100$ (the latter choice thus completely fixing our units). Note that because of the Dirichlet boundary conditions that we used to implement our discretized model, we will only be able to trust our results for maximum integration times $T$ of the order of the size of the lattice $L$ (light crossing time) beyond which spurious reflections are expected to spoil the predictivity of the numerical analysis.

Notice also that because of the matrix nature of the $Z$ variable, the computational complexity of this numerical problem increases as $N^2\propto 1/a^2$ or in other words as the inverse square of the spatial resolution.  

\begin{figure*}[t]
\subfloat[%
Spatial profile of the breather ($\phi_i$ as a function of the lattice position $i$) in the absence of evaporation {\it i.e.} when $\lambda=0$. The snapshots are taken every one sixteenth of a period and later times correspond to lighter shades of gray. 
\label{profileBreatherNonBack}]{%
\includegraphics[width=\columnwidth]{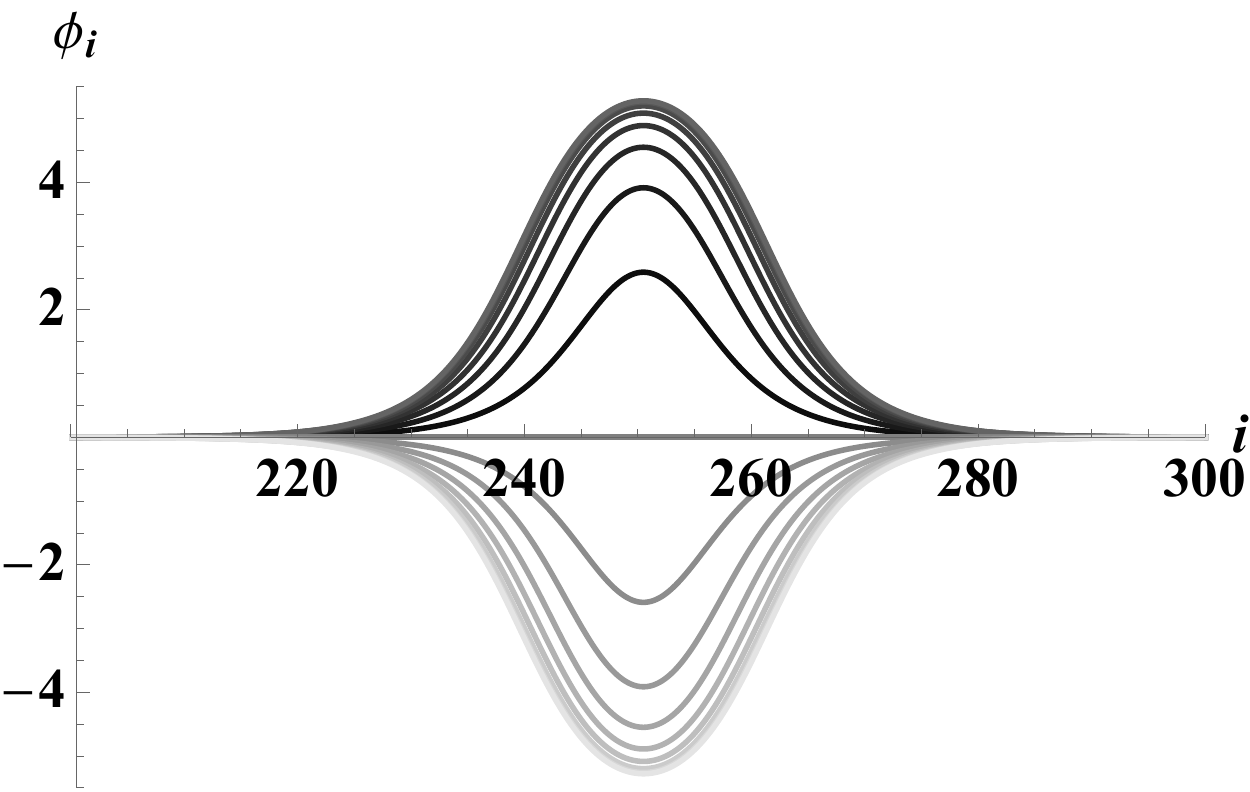}}
\quad
\subfloat[%
Spatial profile of the breather ($\phi_i$ as a function of the lattice position $i$) with evaporation taken into account {\it i.e.} when $\lambda=0.1$. The snapshots are taken every one sixteenth of a period (of the non-evaporating breather) and later times correspond to lighter shades of gray.  
\label{profileBreatherBack}]{%
\includegraphics[width=\columnwidth]{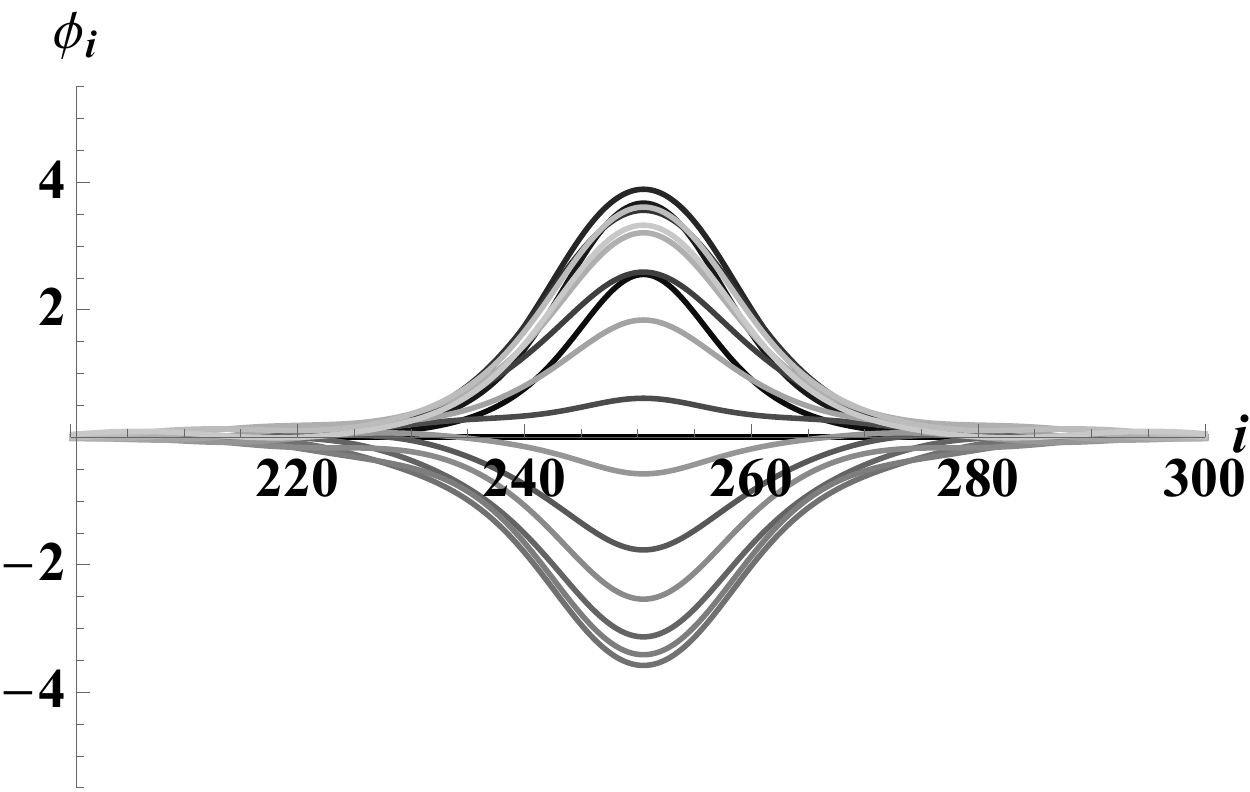}}
\\
\subfloat[%
Breather amplitude {\it i.e.} value of the field at the center of the lattice, as a function of time in the absence of evaporation {\it i.e.} when $\lambda=0$. 
\label{amplitudeBreatherNonBack}]{%
\includegraphics[width=\columnwidth]{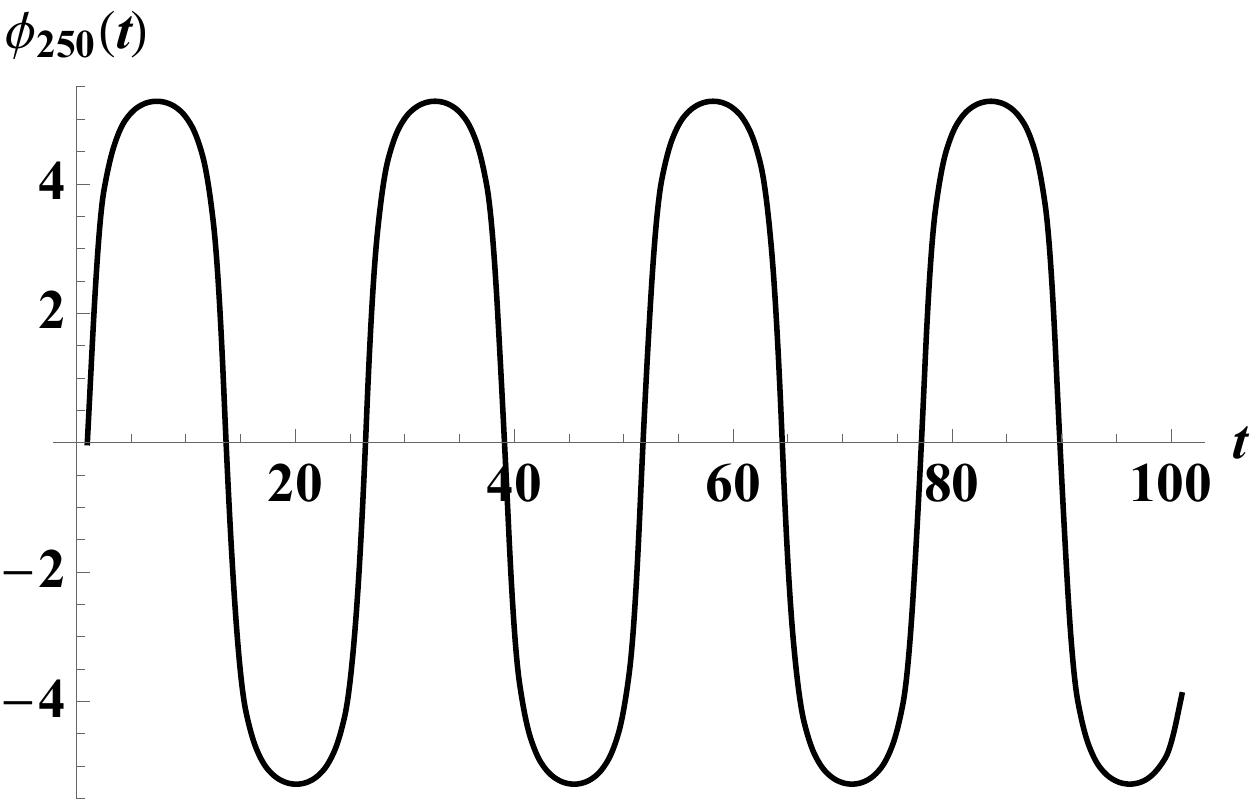}}
\quad
\subfloat[%
Breather amplitude {\it i.e.} value of the field at the center of the lattice, as a function of time with evaporation taken into account {\it i.e.} when $\lambda=0.1$.
\label{amplitudeBreatherBack}]{%
\includegraphics[width=\columnwidth]{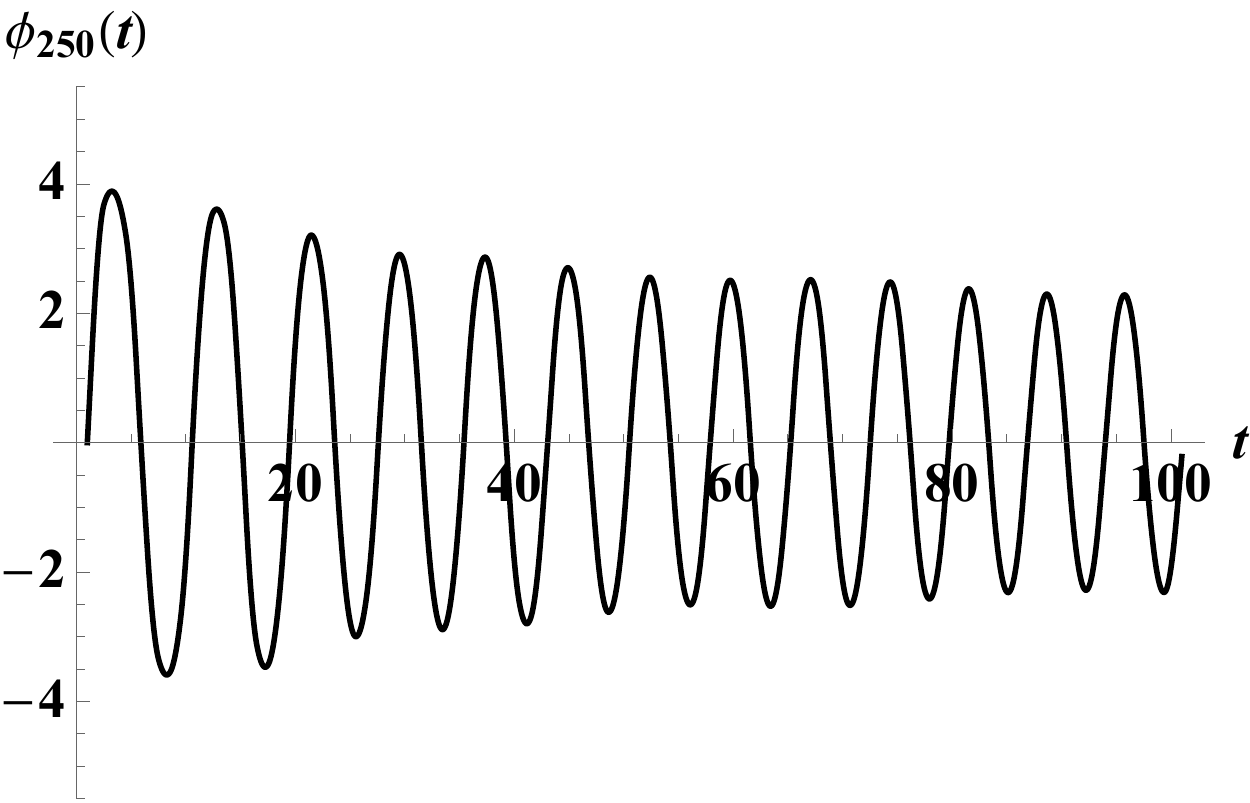}}
\caption{Visualisation of the breather dynamics with and without backreaction. Here $N=500$.}
\end{figure*}
 
Before going on to discuss our numerical results, we need to address the dependence
of our results on the lattice spacing $a$.
Backreaction of the quantum radiation on the classical breather is
encoded in the $Z^*Z$ factor in parentheses
appearing in Eq. \eqref{eqback}. The factor diverges logarithmically as $a \to 0$ 
(for fixed $L = N a$) since, at the initial time we find
\be
\frac{1}{a^2}\sum_{j=1}^NZ^*_{ij}Z_{ij} \sim \frac{1}{2\pi} \ln N 
\sim \frac{1}{2\pi} \ln (\mu L)
\ee
where $\mu \equiv 1/a$ is the energy scale corresponding to the lattice cutoff. 
If we now rescale 
$\mu \to \zeta \mu$, the $Z^*Z$ factor shifts by $\ln(\zeta)/(2\pi)$. In \eqref{eqback},
a shift in the $Z^*Z$ factor is equivalent to a shift in the mass term of $\phi$. Hence to
compare results at different lattice spacings we should ensure that the classical
potential $V(\phi)$ is correspondingly adjusted to obtain the same physical
mass for $\phi$. In other words, if we rescale $\mu \to \zeta \mu$, we should
also change the potential $V(\phi) \to V(\phi ) - \lambda \ln(\zeta) \phi^2/(4\pi)$, and 
then \eqref{eqback} is invariant under the rescaling (except for discretization effects
in the Laplacian term).
 
Now that we know how $V(\phi)$ changes with a rescaling of the lattice spacing
$a$, we need a physical input that tells us the potential for some particular value
of the lattice spacing, call it $a_*$. 
Given that $a$ plays the role of the spatial resolution, we must require that this is enough to resolve the smallest physical object in the simulation box, the breather, whose size is of order 1 (in units where $m_{\phi}=1$). For our purposes it will suffice to take $a_* \approx 0.2$ (which corresponds to a lattice with $N_*=500$ points)
and assume that $V(\phi )$ is the classical sine-Gordon 
potential at this resolution.

The equations of motion \eqref{eqback} and \eqref{eqZ} with initial conditions \eqref{initphiexp},\eqref{initphidotexp}, \eqref{initZexp}, \eqref{initZdotexp} 
can now be solved numerically using explicit methods in Fortran and also in Mathematica (for smaller values of
$N$). As noted above, the computational costs scale as $N^2$ and we are effectively limited to $N \le 1000$.

\section{Results}
\label{results}

As discussed in the introduction, we expect the classical breather field $\phi$ to evaporate under the effect of the coupling to the quantum radiation field $\psi$. This can be explicitly seen in Figs.~\ref{profileBreatherNonBack} and \ref{profileBreatherBack} where we plot snapshots of the breather spatial profile with and without quantum backreaction taken into account: indeed the evaporating breather oscillates at a higher frequency and with decreasing amplitude as compared to the non-evaporating one. The amplitude plots of Figs.~\ref{amplitudeBreatherNonBack} and \ref{amplitudeBreatherBack} confirm these heuristic observations.

We are in particular interested in the breather's decay rate and in how this rate changes with the physical coupling constant $\lambda$. As mentioned previously, we
would also like to check if our renormalization scheme leads to similar evaporation rates when we change $N$.

\begin{figure}[t]
\includegraphics[width=\columnwidth]{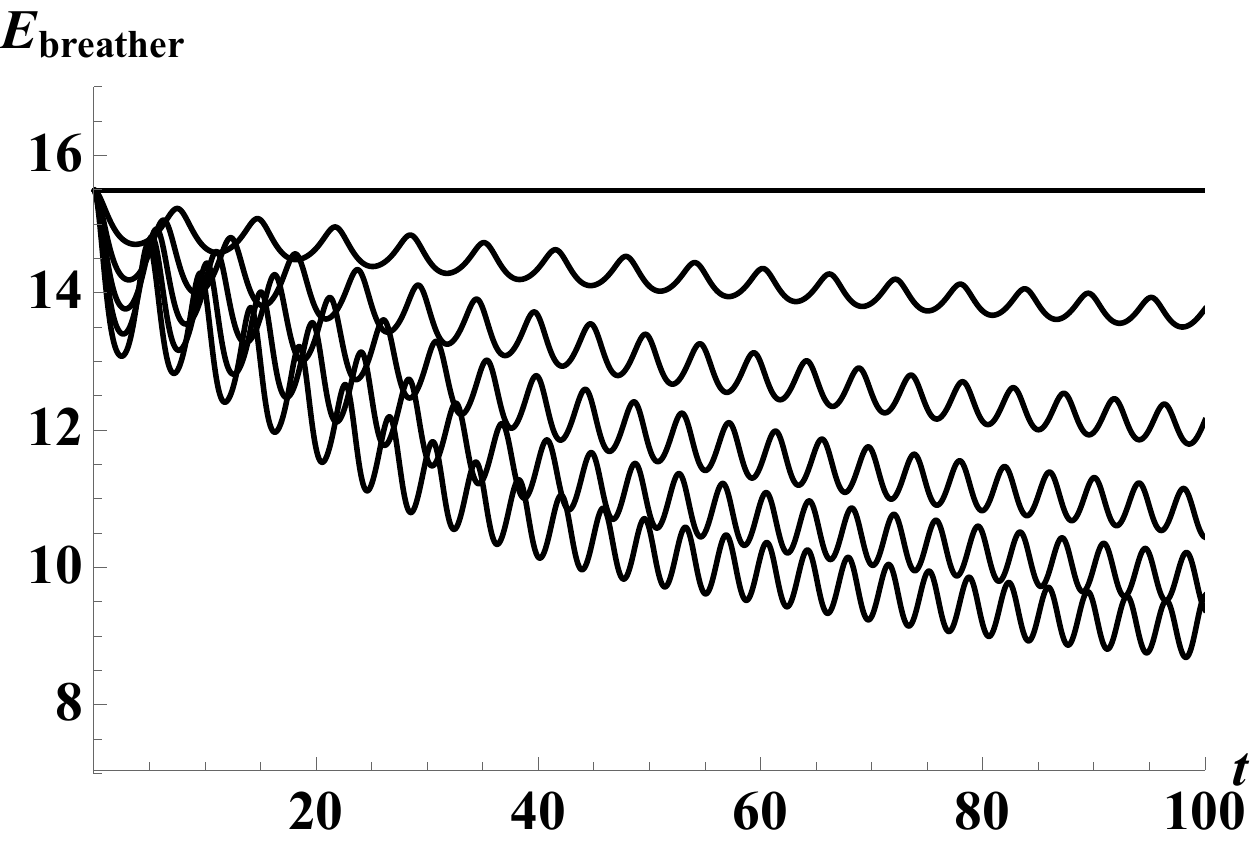}
\caption{Breather energy as a function of time for different values of the physical coupling constant $\lambda$. The top (flat) curve corresponds to $\lambda=0$ while the decaying oscillatory curves correspond, from top to bottom, to $\lambda=$0.02, 0.04, 0.06, 0.08 and 0.1. Here $N=500$.}
\label{energybreatherlambda}
\end{figure}

\begin{figure}[ht]
\includegraphics[width=\columnwidth]{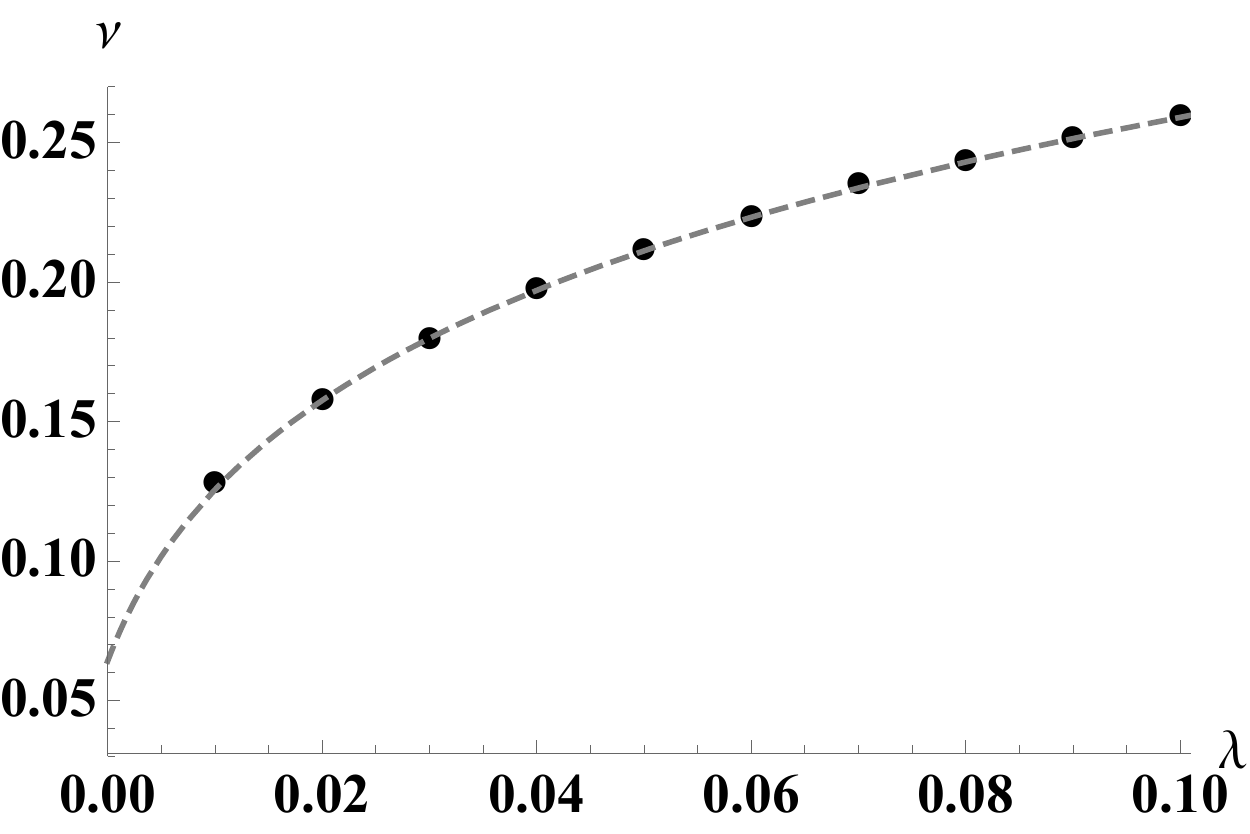}
\caption{Average frequency of oscillation $\nu$ of the breather energy as a function of $\lambda$.
(The oscillation frequency of the breather amplitude is half that of the energy because $E$ is quadratic
in $\phi$.)
The dashed curve represents the best fit power law model: $\nu\approx -0.44 + 0.90(0.01+\lambda)^{0.11}$. Here $N=500$.}
\label{freqsBreather}
\end{figure}

\begin{figure}[ht]
\includegraphics[width=\columnwidth]{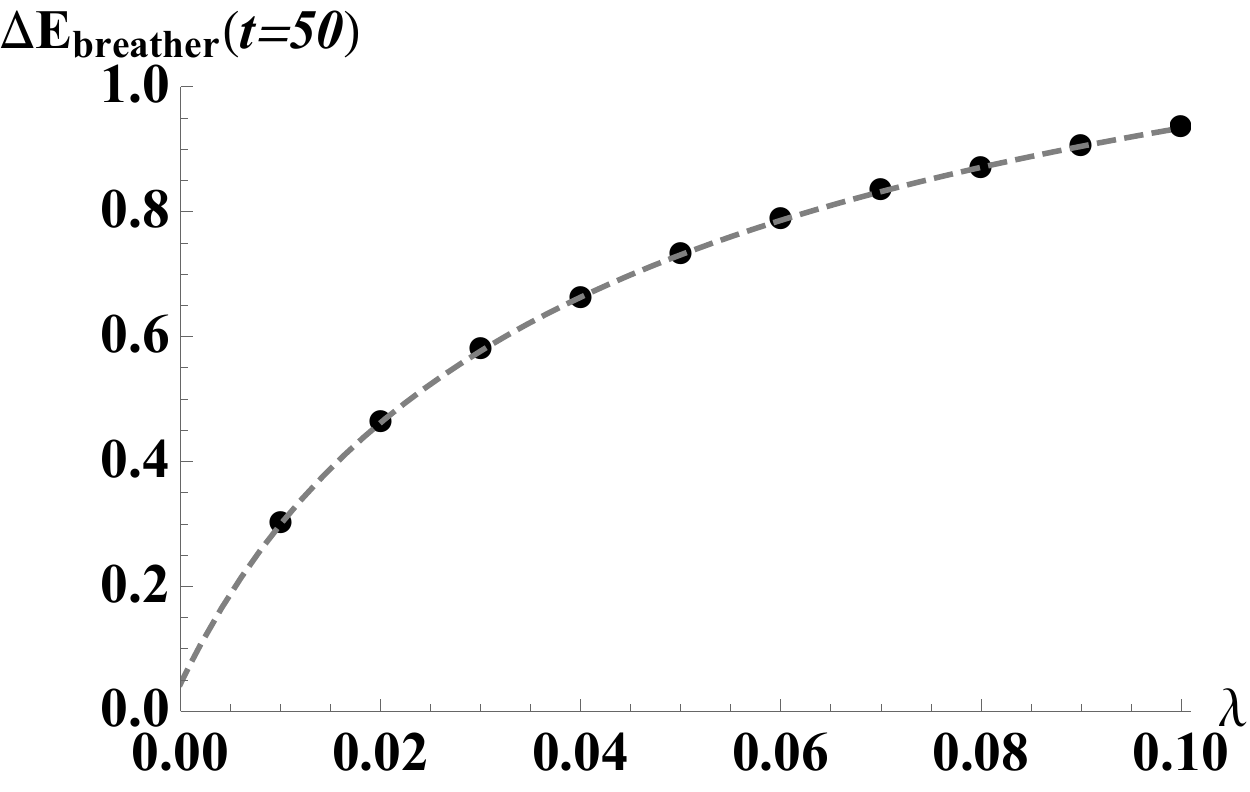}
\caption{Amplitude of oscillation of the breather energy (around time $t=50$) as a function of $\lambda$. The dashed curve represents the best fit power law model: $\Delta E_{\rm breather}(t=50)\approx 1.55 - 0.20/(0.03+\lambda)^{0.56}$. Here $N=500$.}
\label{ampsEBreather}
\end{figure}

In Fig.~\ref{energybreatherlambda} we plot the energy of the breather
\ba
E_{\rm breather}&=&\int dx\left(\frac{1}{2}\dot{\phi}^2+\frac{1}{2}\phi'^2
+m_\phi^2(1-\cos\phi)\right)\nn\\
&\approx& a\sum_{i=1}^N \Biggr[\frac{1}{2}\dot{\phi_i}^2
-\frac{1}{2a^2} \phi_i(\phi_{i-1}-2\phi_i+\phi_{i+1})\nn\\
&&\ \quad\qquad\qquad+m_\phi^2(1-\cos\phi_i)\Biggr] \,,
\ea
as a function
of time for several different values of $\lambda$, keeping all other parameters
fixed. With $\lambda=0$, the breather does not decay, as expected. For non-vanishing $\lambda$, the breather decays while also undergoing small oscillations. As $\lambda$
becomes larger, the breather decays faster and the amplitude and frequency of the oscillations grow (which can readily be seen on Figs.~\ref{freqsBreather} and \ref{ampsEBreather}).
 
\begin{figure}[ht]
\includegraphics[width=\columnwidth]{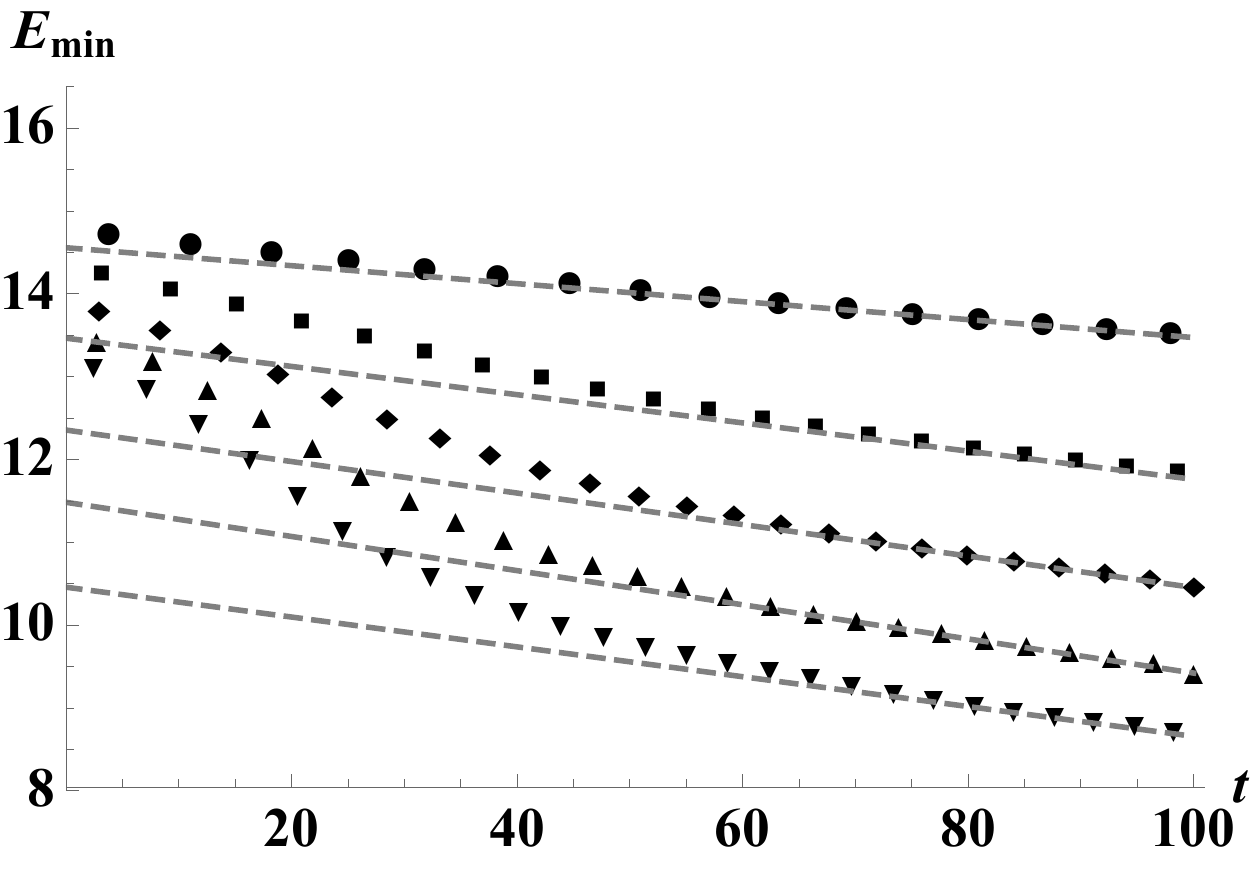}
\caption{Local minima of the breather energy as a function of time for $\lambda=$0.02, 0.04, 0.06, 0.08 and 0.1 (corresponding to circle, square, diamond, upper and lower pointing triangles respectively). Also shown are the respective linear fits for late times with slopes of absolute values 0.011, 0.017, 0.019, 0.021, 0.018. Here $N=500$.}
\label{fitEmin}
\end{figure}

For a more quantitative result we could furthermore find a power law fit of the upper and lower envelopes bounding 
this oscillatory behavior to estimate the breather lifetime and understand its dependence on $\lambda$ in particular. 
Here we only show a linear fit of the large time behavior of the lower envelope in Fig.~\ref{fitEmin} for different values of 
$\lambda$. With the exception of the $\lambda=0.1$ case where the fit is poorer, the slopes increase in absolute 
value with increasing $\lambda$ which is consistent with the intuitive idea that the breather lifetime decreases as the 
coupling becomes stronger.
 
\begin{figure*}[t]
\subfloat[%
Breather energy as a function of time for different values of $N$ 
without mass renormalization. $N$ varies from 200 (lightest shade of grey) to 
1000 (darkest shade of grey) in increments of 100.
\label{energybreathernonre}]{%
\includegraphics[width=\columnwidth]{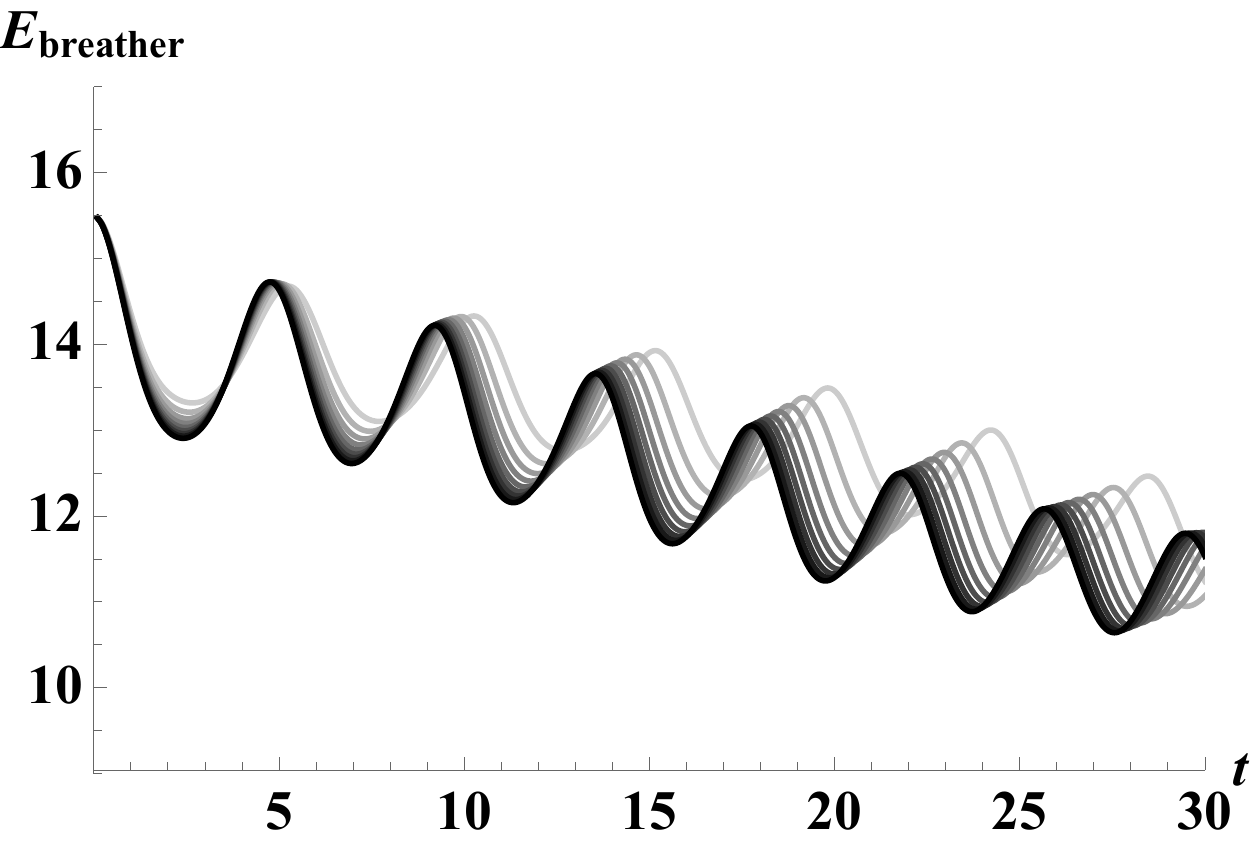}}
\quad
\subfloat[%
Breather energy as a function of time for different values of $N$ 
with mass renormalization. $N$ varies from 200 (lightest shade of grey) to 
1000 (darkest shade of grey) in increments of 100.
\label{energybreatherre}]{%
\includegraphics[width=\columnwidth]{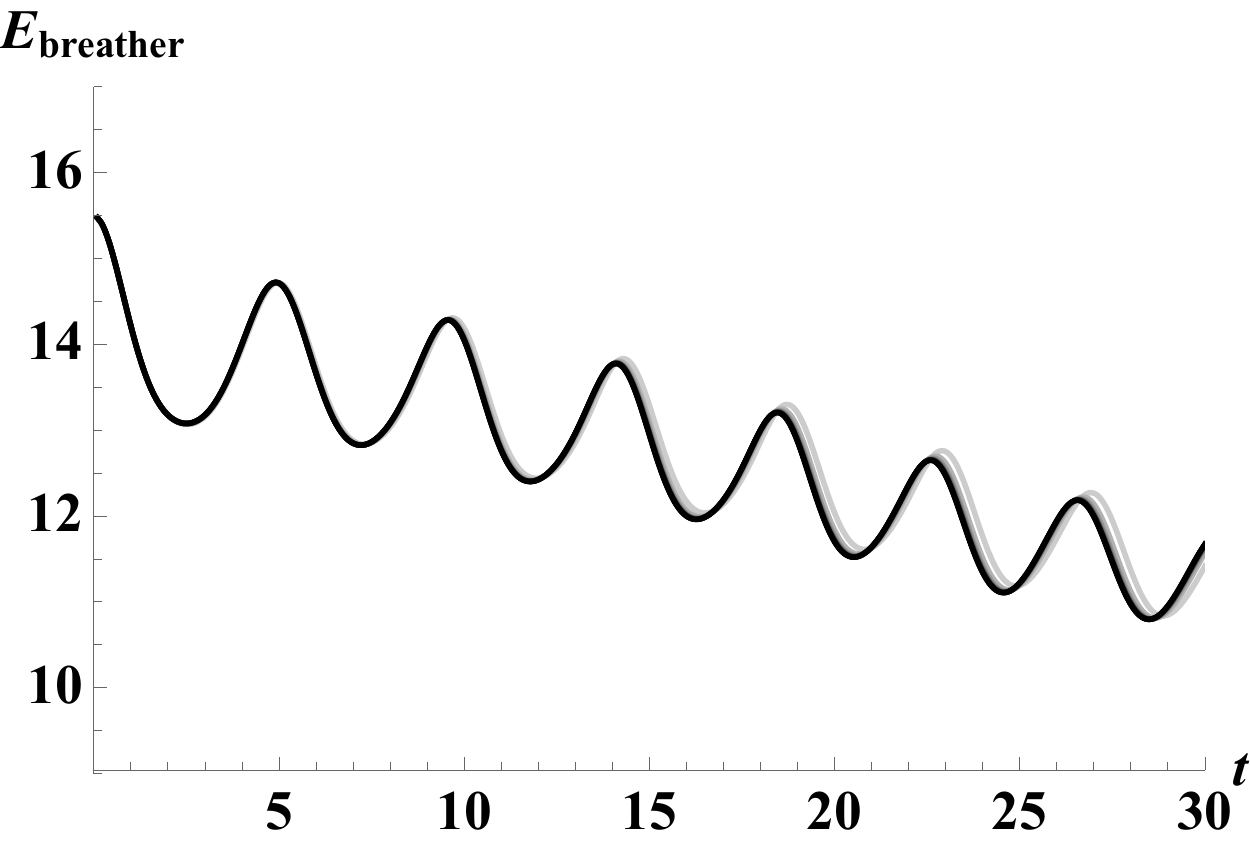}}
\caption{Sensitivity of the breather energy on $N$ with and without coupling constant renormalization.}
\end{figure*}

Finally we check that our renormalization prescription is the correct one for dealing with 
the small $a$ limit. To do this we plot the breather energy for different values of $N$ 
for the same $L$ and $\lambda$ with and without mass 
renormalization (Figs.~\ref{energybreathernonre} and~\ref{energybreatherre}).
It is clear from these two plots that the dependence on the effective description is dramatically mitigated through the use of our renormalization prescription: the obvious drift in Fig.~\ref{energybreathernonre} is greatly reduced when we
 shift the value of the bare mass squared by $\lambda \ln(a/a_*)/(2\pi)$ 
in order for $m_\phi$ to correspond to the physical mass of the sine-Gordon field.
The small residual drift can be attributed to finite $N$ effects {\it i.e.} contributions to the renormalization
that are sub-leading in $N$ and numerical errors due to the coarseness of the lattice. In the $N\to \infty$
limit both these contributions will vanish.

These results, especially Figs.~\ref{energybreatherlambda} and \ref{fitEmin}, give clear evidence that the 
breather evaporates by emitting quanta of $\psi$. 
In our setup, this comes together with oscillations in the breather energy.
In order to understand the origin of these oscillations, note that
the vacuum fluctuations of $\psi$ give rise to a nonzero $\langle \psi^2 \rangle$, which is affected by the boundary 
conditions imposed on our simulation box of finite length $L$ as in the Casimir effect.

It is natural to ask how the initial quantum fluctuations of $\psi$, as measured by $\langle \psi^2 \rangle_{t=0}$, affect the breather evolution. This is very simple to answer by considering the SG model  perturbed by a constant mass term, 
$\Box\phi+\sin\phi+m_{\rm eff}^2\,\phi=0$, with $m_{\rm eff}^2=\lambda \langle \psi^2 \rangle_{t=0}$.
The energy stored in the breather in this model (without including the $m_{\rm eff}^2$ perturbation term) exhibits 
very similar behaviour to that in Figs.~\ref{energybreatherlambda} and \ref{fitEmin}, including the oscillations with 
a definite $\lambda-$dependent period. More specifically, taking $m_{\rm eff}^2 =  \lambda \sum Z^*Z/a^2$ evaluated at the center of the box (and at $t=0$), the plot in Fig.~\eqref{freqsBreather} is reproduced 
with better than 10\% precision. Note that, the frequency is seen to approach twice the breather frequency at $\lambda\ll1$, which is readily understood because the perturbation of the SG model at lowest order in 
$\lambda$ is proportional to $\phi^2$. 

This suggests that the oscillatory behaviour in the breather energy can be interpreted as originating from the 
quantum vacuum fluctuations of $\psi$. On top of this, there is the quantum creation of $\psi$ particles which 
leads to the decrease of the average breather energy over time. Thus the CQC scores a double goal, 
capturing the two kinds of quantum effects at once -- as it should.

\section{Conclusions}
\label{discussion}

In the previous section we have studied the decay of the breather numerically via the CQC. Backreaction was fully taken into account. A legitimate question that might arise is whether the predictions of the CQC actually correspond to the full quantum solution {\it i.e.} where both breather and radiation field are treated quantumly. It turns out that in the fixed background approximation the $1/a^2\sum_{j=1}^NZ^*_{ij}Z_{ij}$ term responsible for the backreaction is actually exactly equal to the vacuum expectation value $\langle 0|\psi_i^2|0\rangle$. Therefore the solution of the dynamical system discussed in this paper can be seen to correspond to the limit of a semi-classical iterative procedure where one starts with the classical breather solution, computes the quantum radiation in this background, uses it to calculate the first semi-classical correction to the background and repeats this process {\it ad infinitum}. To our knowledge this is the first time such a numerical calculation is carried out in full. Usually, the first iteration of this semi-classical procedure is carried out and claimed to give accurate results. However this can lead to relative errors of order $100\%$ as discussed in Ref.~\cite{Vachaspati:2018llo} in the context of a $0+1$ dimensional toy model. 
Previous work on the quantum decay of oscillons~\cite{Hertzberg:2010yz} 
used methods analogous to the CQC to calculate
the radiation rate on a fixed oscillon background. However 
the backreaction on the oscillon background was not computed.

The method outlined in this paper can be very powerful in computing quantum backreaction on classical 
backgrounds but it is important to be aware of its limitations in order to successfully apply it to other field 
theory models. First of all, it only yields accurate results in the limit where the background can actually be 
treated classically. This is only true in the limit where the pure sine-Gordon part of the action evaluated for 
the breather solution dominates the quantum radiation as well as the interaction parts (this can be seen by 
temporarily restoring the reduced Planck constant $\hbar$). We see that this requirement amounts 
to~\cite{Coleman:1985rnk}
\be
E_{\rm breather}\sim16m_\phi\sqrt{1-\frac{\omega^2}{m_\phi^2}}\gg E_Z - \frac{1}{2}\text{Tr}\;\Omega_0\,,
\ee
where $E_Z=\sum_{i,j,k=1}^N [ {\dot Z}^*_{ij}\delta_{ik} {\dot Z}_{kj}/2a + 
Z^*_{ij} \Omega^2_{ik} Z_{kj}/2a ]$ and the subtraction in the last term is meant to remove the vacuum energy contribution (which diverges quadratically with $N$). For our choice of parameters and since total energy is conserved, Fig.~\ref{energybreatherlambda} shows that this condition will hold up to times $T\sim L$ as long as $\lambda\ll 1$.

Second, our method also has intrinsic numerical limitations chief among which is its quadratic computational complexity in the size of the lattice. This renders high resolution computations intractable and is a source of numerical error which manifests itself in what could be termed finite $N$ artifacts (such as the residual $N$-dependent drift in the energy of the breather even after coupling constant renormalization, or the imperfect conservation of total energy). Another source of inaccuracy are the perfectly reflecting boundary conditions chosen in the discretization of the continuous problem which set an upper bound on the meaningful integration time. 

However these limitations also suggest new avenues for improvement. In particular parallelizing the code would allow for larger lattices, both potentially increasing the resolution and pushing the spatial boundaries further so as to allow for larger integration times. One could also implement higher order discretization schemes to increase accuracy, or absorbing boundary conditions to reduce spurious reflections. 

Finally, it should be mentioned that this technique should be readily applicable to more complicated field theory scenarios such as backreaction of quantum radiation on a gravitationally collapsing background or Schwinger pair creation (recall however that the quantum field cannot have self-interactions). However, gravitational scenarios will necessarily involve more intricate renormalization schemes in particular when dealing with vacuum energy divergences. Another interesting question is whether one can use the CQC to obtain the quantization of the breather spectrum in the full quantum SG theory \cite{Coleman:1985rnk}. By separating the SG field into a classical breather plus quantum fluctuations and treating the latter via CQC one may perhaps reproduce this classic result.

We have thus solved for the quantum evaporation of classical breathers by applying the CQC. This is 
the first application of the CQC to a field theory problem that has not been solved by 
traditional methods. The CQC relies on the classicality of the background variable and 
if the background is quantum there will be deviations from the CQC as seen in the quantum mechanical
example in~\cite{Vachaspati:2018llo}. Thus it will still be useful to solve the breather evaporation 
problem in full quantum field theory in order to compare to the CQC result and to understand the
method's limitations.
However, since the breather is non-perturbative, we expect that a quantum field theory treatment
will require new techniques and/or a lattice implementation. In the context of gravity, the CQC
may be the only hope to solve important problems such as black hole evaporation because
we do not yet have a quantum theory of gravity.\\

\acknowledgements
TV is supported by the U.S. Department of Energy, 
Office of High Energy Physics, under Award No. DE-SC0019470 at Arizona State 
University and GZ is supported by John Templeton Foundation grant 60253.\\
JO and OP acknowledge support by the Spanish Ministry MEC under grant FPA2017-88915-P, the Severo Ochoa excellence program of MINECO (grant SEV-2016-0588), as well as by the Generalitat de Catalunya under grant 2017SGR1069.
We also acknowledge the support of the Centro de Ciencias de Benasque Pedro Pascual for hosting the workshop `Probes of BSM - from the Big Bang to the LHC' , where this project was initiated.

\bibstyle{aps}
\bibliography{cqcBreather}

\end{document}